\documentclass[runningheads]{llncs}
\usepackage[T1]{fontenc}
\usepackage{graphicx,verbatim}
\usepackage{amsmath,amssymb,bm}
\usepackage{booktabs}
\usepackage{multirow}
\usepackage[table]{xcolor}
\usepackage{hyperref}
\hypersetup{
  colorlinks=false,            
  citebordercolor={0 1 0},     
  linkbordercolor={0 1 0},     
  urlbordercolor={0 1 0}
}

\begin{document}


\title{A Differentiable Brain Tumor Mass Effect Solver}

\author{Marco L. Wittrich \inst{1,2}\and
Michal Balcerak \inst{4} \and
Lucas Zimmer \inst{1} \and
Ivan Ezhov \inst{3} \and
Björn Menze \inst{4} \and
Marie-Christin Metz \inst{1} \and
Nicole Tueni \inst{5} \and
Benedikt Wiestler \inst{1,2} \and
Jonas Weidner \inst{1,2}
}
\authorrunning{Wittrich et al.}
\institute{
AI for Image-Guided Diagnosis and Therapy, TUM, Munich, Germany \and
Munich Center for Machine Learning (MCML), Munich, Germany \and
Chair for AI in Healthcare and Medicine, TUM, Munich, Germany\and
University of Zurich, Switzerland\and
Institute of Continuum Mechanics and Biomechanics, Friedrich-Alexander-Universität Erlangen-Nürnberg, Erlangen, Germany\\
\email{\{m.wittrich, j.weidner\}@tum.de} 
}

\maketitle

\begin{abstract}
A growing brain tumor mechanically deforms the surrounding healthy brain through mass effect. This compression and displacement of tissue affects both patient prognosis and treatment planning, so biophysical models of tumor growth must account for this deformation.
Established mass-effect models are partial differential equation (PDE) solvers that couple growth to elasticity and calibrate parameters per patient, at substantial implementation, runtime, and calibration cost. We ask how much of this deformation a fully differentiable energy-based model can recover, and introduce a potential-spring model: a regular grid over the brain is driven outward by a tumor-concentration potential and held by an elastic spring lattice with tissue-specific stiffness. It shares the tumor-gradient driving force of continuum elasticity but uses a single global parameter and no per-patient inversion, minimized end-to-end by gradient descent.
We benchmark it against established solvers by applying each displacement field to a healthy atlas and measuring tissue overlap with the patient's own segmentation on 134 BraTS patients.
Our model yields the largest overall local overlap improvement at up to 100× shorter computation time, establishing a lightweight, differentiable solver for tumor mass effect.
Because our model obeys mechanical laws and reflects the tumor-induced tissue loading, it remains physically plausible, a prerequisite for clinical use.

\end{abstract}

\section{Introduction}
Glioblastoma is the most common and aggressive primary brain tumor, and predicting where and how it grows is central to defining radiotherapy targets. Tumor-growth models based on the reaction-diffusion equations (Fisher - Kolmogorov) predict a tumor \emph{concentration} field, but the growing tumor also \emph{mechanically} deforms the surrounding healthy tissue.
This so-called mass effect is itself clinically relevant and feeds back on the pathology. Modeling the induced tissue deformation is crucial for downstream applications. Existing mass effect models are biophysical PDE solvers: they couple tumor growth to continuum elasticity and either invert the mechanical parameters per patient \cite{scheufele2019,lipkova2022modelling} or solve the linear-elastic equilibrium forward with fixed parameters \cite{ezhov2023learn}. Those solvers achieve high fidelity but entail long runtime and calibration burdens. Furthermore, those solvers are non-differentiable, which makes them infeasible for gradient-based optimization.

Another line of work avoids explicit forward simulation by directly optimizing a discrete loss for tumor growth, including mass effect \cite{balcerak2024physics,Balcerak2025GliODIL}. In this setting, the PDEs serve as soft regularizers alongside an image-based data term, rather than as hard physical constraints. This makes the optimization flexible, but the resulting solution is not guaranteed to satisfy the coupled growth and mechanics equations exactly.

We propose a rapid, fully differentiable energy-based model that acts on a pre-calculated tumor cell concentration. We make three main contributions: 
(i) We introduce a new \emph{potential-spring} based biophysical model of tumor mass effect: a differentiable grid energy that uses the same tumor-gradient driving force and tissue-specific stiffness (from brain elastography) as continuum elasticity, but a single global parameter and no per-patient inversion, and is minimized end-to-end by gradient descent.
\footnote{The code is available on \url{https://github.com/Codorado/potential-spring-model}.}
(ii) We benchmark it against two PDE solvers, mass effect solver of GLIA (Glioblastoma Image Analysis) suite (coupled growth with per-patient inversion) \cite{subramanian2022ensemble} and a forward linear-elasticity solver (FLE) \cite{ezhov2023learn}, together with an ANTs SyN deformable-registration \emph{upper bound}, on 136 BraTS2021 \cite{baid2021rsna} patients with a ventricle-specific analysis. 
(iii) We show that the models are \emph{complementary}: our proposed model leads on the parenchyma and GLIA on the ventricles, established with a tissue-overlap validation (per-tissue Dice and reclassification in peritumoral bands).

\begin{figure}[t]
\centering
\includegraphics[width=\textwidth]{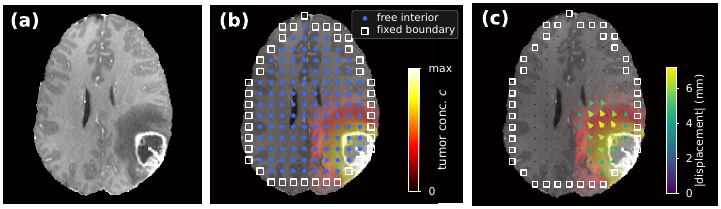}
\caption{\textbf{The potential-spring model (PSM).} 
(a)~Axial T1c.
(b)~T1c overlaid with the tumor-concentration potential $c$ and the \emph{undeformed} spring grid (free interior nodes vs.\ the fixed boundary shell).
(c)~The converged grid (minimizer of Eq.~\eqref{eq:spring}) shown as displacement vectors
(color~$=$~magnitude in mm). Nodes are driven outward, down the tumor-concentration gradient, and balanced by the elastic restoring force of the spring lattice, which couples each node to its 26 neighbors in the $3\times3\times3$ cube. Displacement is largest where the concentration gradient is steepest.}
\label{fig:concept}
\end{figure}

\section{Methods}
We present a differentiable energy-based mass-spring model of tumor mass effect and implement literature baselines for comparison.
\subsection{Proposed potential-spring mass effect model}
The Potential Spring Model (PSM) uses a regular grid that is laid over the brain and we treat each node as a point mass linked to its neighbors by springs. Interior nodes are free to move while a shell of nodes on the brain boundary is held fixed, imposing a rigid-skull (Dirichlet) boundary condition (see Figure~\ref{fig:concept}. The tumor acts as a potential that pushes nodes outward down its concentration gradient, while the spring lattice penalizes deviations of inter-node distances.
Each interior node is connected to the 26 neighbors of its surrounding 3x3x3 cube.
We minimize the energy
\begin{equation}
E(\mathbf{u}) \;=\;
\underbrace{\alpha \!\!\sum_{i\in\mathcal{F}}\! c(\mathbf{x}_i+\mathbf{u}_i)}_{\text{tumor potential}}
\;+\;
\underbrace{\beta \!\!\sum_{(i,j)\in\mathcal{E}}\! k_{ij}\,
\Big(\big\lVert (\mathbf{x}_i+\mathbf{u}_i)-(\mathbf{x}_j+\mathbf{u}_j)\big\rVert - r^{0}_{ij}\Big)^{2}}_{\text{tissue-weighted spring lattice}},
\label{eq:spring}
\end{equation}
 over the node displacements
$\{\mathbf{u}_i\}$, where $c$ is the tumor concentration (potential),  $\mathcal{F}$ the free intra-brain nodes, $\mathcal{E}$ the spring edges, and $r^{0}_{ij}$ the rest length of edge $(i,j)$.
The edge stiffness $k_{ij}=\tfrac12(k_i+k_j)$ averages tissue-specific node stiffnesses from region-specific brain elasticity (gray matter $2.7\times$ stiffer than white matter \cite{hinrichsen2023inverse}, CSF $10\times$ softer (empirically chosen)).

The potential term contributes a body force $-\alpha\,\nabla c$, the same tumor-gradient forcing as the continuum solvers (cf.\ Eq.~\eqref{eq:glia}).
The tissue-weighted lattice is in fact a mass-spring discretization of the linear-elastic operator: the face/edge/body-diagonal connectivity renders the discrete stiffness approximately isotropic, and the per-tissue $k_{ij}$ play the role of the spatially variable Lam\'e coefficients $\lambda(\mathbf{x}),\mu(\mathbf{x})$ of Eq.~\eqref{eq:glia}.
Equation~\eqref{eq:spring} is thus a discrete energy form of the same physics the PDE solvers integrate, with a single global force scale $\alpha$ in place of per-patient inversion.

We minimize $E(\mathbf{u})$ over the free-node displacements by gradient descent, with $\nabla_{\mathbf{u}}E$ from JAX automatic differentiation, and interpolate the node displacements to the voxel grid.
Each patient needs a single forward run and no inversion, and because the solve is differentiable it can be embedded in larger
gradient-based pipelines. The one free parameter, the force/stiffness ratio $\alpha$ ($\beta$ fixed), is set to $\alpha=1000$ determined by grid search and maximizing net reclassification \ref{sec:model_evaluation} while keeping the deformation invertible. We measure invertibility by the Jacobian $J=\det(\mathbf{I}+\nabla\mathbf{u})$: the folded fraction (voxels with $J\le0$, where the warp is non-invertible) must stay near zero.

\subsection{Established biophysical baselines}
We benchmark PSM against two biophysical PDE solvers, FLE \cite{ezhov2023learn} (forward linear elasticity, no inversion) and GLIA \cite{subramanian2022ensemble} (per-patient PDE inversion), and an image-based registration ceiling (ANTs SyN), whose objective is the same anatomical overlap as our evaluation. FLE takes the same tumor concentration as PSM, while GLIA reconstructs its own.

\paragraph{Biophysical PDE solvers (GLIA \cite{subramanian2022ensemble} and FLE \cite{ezhov2023learn}).}
Both are built on a reaction-diffusion tumor coupled to linear elasticity. The tumor evolves as
\begin{equation}
\partial_t c \;=\; \nabla\!\cdot(\mathbf{D}\,\nabla c) \;+\; \rho\,c\,(1-c),
\label{eq:rd}
\end{equation}
and quasi-static mechanical equilibrium is driven by the tumor gradient,
\begin{equation}
\nabla\!\cdot\bm{\sigma}(\mathbf{u}) \;=\; \gamma\,\nabla c,
\qquad
\bm{\sigma} = \lambda(\mathbf{x})\,\mathrm{tr}(\bm{\varepsilon})\,\mathbf{I} + 2\mu(\mathbf{x})\,\bm{\varepsilon},
\quad
\bm{\varepsilon}=\tfrac12\!\left(\nabla\mathbf{u}+\nabla\mathbf{u}^{\top}\right).
\label{eq:glia}
\end{equation}
Here $c\in[0,1]$ is the tumor cell concentration and $\lambda(\mathbf{x}),\mu(\mathbf{x})$ are per-tissue Lam\'e coefficients (stiff parenchyma, soft CSF, near-rigid skull). The $+\gamma\nabla c$ forcing is the gradient of a tumor pressure $p=\gamma c$, so $\gamma{=}0$ gives $\mathbf{u}{=}0$. Its sign matches the GLIA source and \cite{subramanian2020multiatlas}.
GLIA solves Eq.~\eqref{eq:glia}, which is the per-step-equilibrium while \eqref{eq:rd} evolves the tumor in time.
FLE instead only solves \eqref{eq:glia} for a fixed tumor concentration. 

GLIA solves an inverse problem for each patient via PDE-constrained optimization.
It reconstructs the tumor's initial location and inverts the proliferation, diffusion and mass effect parameters ($\rho$, $\kappa$, $\gamma$) by fitting the simulation to the patient's tumor segmentation data. Because GLIA explicitly models ventricle compression, it requires a healthy tissue atlas that contains the ventricles. We run it in single-atlas mode on the shared 4-label atlas (CSF/GM/WM/VT, the same atlas every method deforms), affine-registered into each patient's SRI frame, with the inversion on a $128^3$ grid.

FLE is the Learn-Morph-Infer brain tumor solver of Ezhov et al.~\cite{ezhov2023learn} run forward (no inversion) as a single-shot static solve of the fixed tumor (the direct analog of our spring solve). The Lam\'e and forcing parameters were fixed and taken from the literature~\cite{subramanian2019simulation}. Its global $\gamma$ scale is set at the maximum-net-before-folding point (displacement is linear in $\gamma$), optimized on a subset of 40 samples.

Relative to both, our model uses the same $\gamma\nabla c$ (here $-\alpha\nabla c$) driving force but a discrete, statically decoupled mass-spring lattice with one global $\alpha$, instead of co-evolved growth and deformation with per-patient inversion.

\paragraph{Deformable registration (ANTs SyN)}
ANTs SyN \cite{avants2009advanced,avants2008symmetric} computes a deformation that warps an atlas T1 image to a patient's T1 image and is invertible in its default configuration. Its discrete realization on the voxel grid can, however, show foldings (see \ref{table:field_metrics}). This deformation field is used as a mass effect proxy. Unlike the biophysical mass effect models, it optimizes anatomical overlap directly. Since we evaluate the models on dice scores, we report the ANTs SyN scores as a ceiling.

\subsection{Data}
We process 136 BraTS2021 patients in the common SRI24 atlas space ($240\times240\times155$; GLIA on $256^3$ through zero-padding; 1\,mm isotropic), each with a tumor segmentation (NCR/ED/ET).
The tumor concentration field $c$ is estimated with a CMA-ES, Fisher-Kolmogorov biophysical growth fit~\cite{weidner2025fiber}. It serves both as the potential in Eq.~\eqref{eq:spring} and as the fixed tumor for FLE, so that the spring-versus-FLE comparison isolates the mass effect model with the tumor input held fixed.
GLIA instead reconstructs its own initial condition from the patient tumor segmentation.
Two GLIA inversions failed, so all methods are evaluated on the shared cohort of 134 patients.

\subsection{Model evaluation}
\label{sec:model_evaluation}
We evaluate the mass effect models on the deformed tissue map and the deformation field characteristics.

\paragraph{Tissue map deformation} Therefore, we evaluate tissue dice between the patient segmentation and the affine registered atlas segmentation first with and second without application of the mass effect predictions.
Each field-based method supplies a displacement field $\mathbf{u}$ in the common frame, applied with the same nearest-neighbor pull-warp to the shared, affine-registered healthy atlas labels $A$,
\begin{equation}
W(\mathbf{x}) \;=\; A\big(\mathbf{x}-\mathbf{u}(\mathbf{x})\big),
\end{equation}
and we score the overlap of the warped atlas tissue $W$ against the patient tissue segmentation $G$.
The ground truth $G$ is a per-patient, brain-confined tissue map segmented from the patient's own T1 (\texttt{deep\_atropos} \cite{tustison2021antsx} with DKT; four classes CSF/GM/WM/VT).
PSM, FLE, and SyN all warp the same shared atlas, whereas GLIA, a coupled growth model whose exported field is only first-order accurate as a single warp, deforms its own copy and is scored from its native deformed atlas.
GLIA is therefore read via the baseline-agnostic $\Delta$Dice.

We evaluate in peritumoral bands (Euclidean shells of 5, 10, 15 and 20\,mm around the tumor) and over the whole brain, always excluding the tumor itself (NCR/ED/ET), where tissue is replaced rather than displaced.
The bands localize where the mass effect acts, which whole-brain Dice would otherwise dilute. 
Our primary metric is the per-tissue Dice of the warped versus the affine atlas, $\Delta\mathrm{Dice}=\mathrm{Dice}(W)-\mathrm{Dice}(A)$ (including the ventricles, GLIA's home metric).
On the disagreement set $\mathcal{N}=\{\mathbf{x}: A\neq G\}$ within a region we additionally report
\begin{equation}
\mathrm{recall}=\frac{|\{\mathbf{x}\in\mathcal{N}: W=G\}|}{|\mathcal{N}|},
\mathrm{net}=\frac{\big|\{A\neq G,\,W=G\}\big|-\big|\{A=G,\,W\neq G\}\big|}{|\mathcal{N}|},
\end{equation}
that is, the fraction of needed tissue corrections achieved (recall) and the net of fixes minus breaks (net reclassification).

\paragraph{Field direction.}
Independently of the segmentation, we also measure how each displacement field is oriented relative to the ANTs SyN deformation field. The magnitude-weighted cosine similarity of a solver's deformation field $\mathbf{u}$ to the SyN field $\mathbf{u}_{\mathrm{SyN}}$ is given by equation \ref{eq:similarity}, with $S=1$ if the fields are parallel and $S=-1$ if antiparallel in every voxel. The voxel contribution to $S$ is weighted with the displacement magnitude. The fields are invariant to the global magnitude scale, allowing a fair comparison between varying field magnitudes.

\begin{equation}
\label{eq:similarity}
S \;=\; \frac{\sum_{\mathbf{x}} \mathbf{u}(\mathbf{x})\cdot\mathbf{u}_{\mathrm{SyN}}(\mathbf{x})}
{\sum_{\mathbf{x}} \lVert\mathbf{u}(\mathbf{x})\rVert\,\lVert\mathbf{u}_{\mathrm{SyN}}(\mathbf{x})\rVert}
\end{equation}

\paragraph{Field plausibility.}
Independently of any segmentation, we summarize each field's physical plausibility over the brain by three measures of the displacement gradient $\nabla\mathbf{u}$.
\emph{Folding} is the fraction of voxels with $\det J\le 0$ (Jacobian $J=\mathbf{I}+\nabla\mathbf{u}$), where the warp is locally non-invertible.
\emph{Smoothness} is the normalized roughness $\langle\lVert\nabla\mathbf{u}\rVert\rangle/
\langle\lVert\mathbf{u}\rVert\rangle$ (in $\mathrm{mm}^{-1}$), the mean displacement-gradient magnitude per unit displacement, where lower values indicate a smoother field. \emph{Vorticity} is the rotational fraction
\begin{equation}
V \;=\; \sqrt{\frac{\sum_{\mathbf{x}} \big\lVert \tfrac12\big(\nabla\mathbf{u}-\nabla\mathbf{u}^{\top}\big)\big\rVert^2}
{\sum_{\mathbf{x}} \lVert \nabla\mathbf{u} \rVert^2}} \in [0,1],
\end{equation}
the share of the gradient energy carried by the antisymmetric (rotational) part of $\nabla\mathbf{u}$.
$V=0$ for a curl-free, purely outward push and $V\approx0.5$ for an isotropic random field. A physically plausible mass-effect field should fold little, vary smoothly, and be strain-dominated (low vorticity).

\section{Experiments and Results}
All scores are computed against the 4-class \texttt{deep\_atropos} ground truth and reported as the mean over subjects $\pm$ SEM. PSM ($\sim$15\,s) and FLE ($\sim$55\,s) are fast single forward solves and the SyN ceiling is a CPU registration ($\sim$32\,s), whereas GLIA's per-patient PDE inversion ($\sim$24\,min) is about $100\times$ slower than PSM for the mass effect solve alone. For PSM and FLE, the tumor concentrations have to be computed in advance, which takes a few seconds, depending on the model \cite{zimmer2025predict,weidner2025lightweight}. GPU methods on a single RTX 8000 (48~GB).

\begin{figure}[t]
\centering
\includegraphics[width=\textwidth]{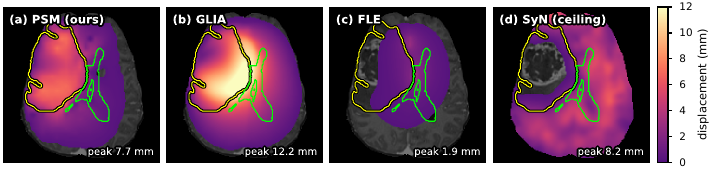}
\caption{\textbf{Qualitative displacement-magnitude comparison} on one large mass effect case
(tumor + edema (yellow) and patient ventricle (green) outlined). The physical solvers a-c show largest deformation within and around the tumor, while SyN (d) as an image-based registration shows displacement outside of the tumor region. PSM produces a broad outward push, GLIA a ventricle-targeted field, FLE a weaker rim-concentrated field, and SyN~(ceiling) a global warp.}
\label{fig:qualitative}
\end{figure}

\begin{figure}[t]
\centering
\includegraphics[width=\textwidth]{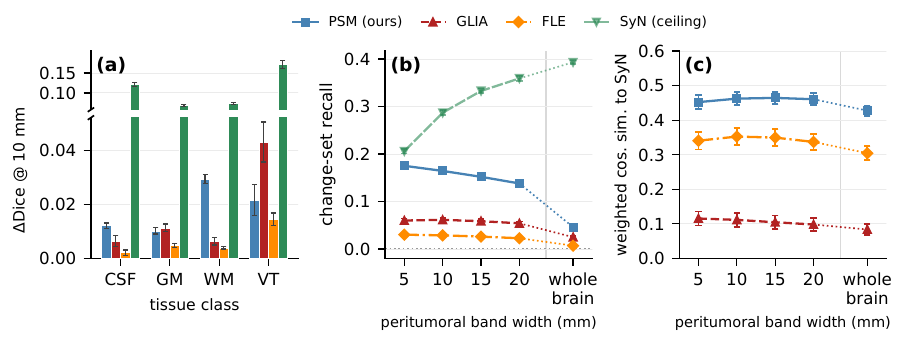}
\caption{\textbf{Quantitative comparison} (mean\,$\pm$\,SEM, $n{=}134$). Our method performs best or equal across all metrics, all distances to the tumor, and most tissues except the ventricles.
(a)~Per-tissue $\Delta$Dice at the 10\,mm band: PSM leads on the white matter and CSF, GLIA on the ventricles (VT), FLE is small throughout, and SyN~(ceiling) is highest (broken axis).
(b)~Change-set recall vs.\ peritumoral band width.
(c)~Weighted cosine similarity of each method's field to the SyN ceiling vs.\ band width.}
\label{fig:quantitative}
\end{figure}

\begin{table}[t]
\centering
\caption{\textbf{Model performance.} Peritumoral (10\,mm) and whole-brain (WB) overlap improvement (tumor-masked, $n{=}134$, mean\,$\pm$\,SEM). $\Delta$Dice is GT-volume-weighted and
averaged over CSF/GM/WM/VT, and net is change-set net reclassification, both in $\times10^{-2}$. SyN is the registration ceiling (upper bound). \textbf{Bold}/\underline{underline} mark the best/second-best biophysical model.}
\label{table:overlap_metrics}
\setlength{\tabcolsep}{3.8pt}
\begin{tabular}{lcccc}
\toprule
\multirow{2}{*}{\textbf{Model}} & \multicolumn{2}{c}{\textbf{$\Delta$Dice} $\uparrow$ [$\times10^{-2}$]} & \multicolumn{2}{c}{\textbf{net} $\uparrow$ [$\times10^{-2}$]} \\
\cmidrule(lr){2-3} \cmidrule(lr){4-5}
& \textbf{10\,mm} & \textbf{WB} & \textbf{10\,mm} & \textbf{WB} \\
\midrule
\rowcolor{gray!20}
  SyN (ceiling) & 8.05\,$\pm$\,0.22 & 11.07\,$\pm$\,0.16 & 17.30\,$\pm$\,0.50 & 23.72\,$\pm$\,0.33 \\
\midrule
  PSM
    & \textbf{2.36}\,$\pm$\,0.14
    & \textbf{0.53}\,$\pm$\,0.04
    & \textbf{4.87}\,$\pm$\,0.29
    & \underline{1.02}\,$\pm$\,0.07 \\
  FLE
    & 0.44\,$\pm$\,0.04
    & 0.06\,$\pm$\,0.01
    & 0.87\,$\pm$\,0.08
    & 0.11\,$\pm$\,0.01 \\
  GLIA
    & \underline{0.88}\,$\pm$\,0.11
    & \underline{0.52}\,$\pm$\,0.05
    & \underline{1.07}\,$\pm$\,0.19
    & \textbf{1.13}\,$\pm$\,0.09 \\
\bottomrule
\end{tabular}
\end{table}

\begin{table}[t]
\centering
\caption{\textbf{Deformation-field characteristics}, (whole-brain,
tumor-masked, mean\,$\pm$\,SEM). Fold $=$ fraction of voxels with $\det J\le 0$;
Smooth.\ $=$ normalized roughness $\langle|\nabla u|\rangle/\langle|u|\rangle$;
Vort.\ $=$ normalized rotational fraction. Folding is ranked (\textbf{bold}/\underline{underline} $=$ lowest/second-lowest
among PSM/FLE/GLIA); smoothness and vorticity are not ranked. SyN is the registration ceiling.}
\label{table:field_metrics}
\setlength{\tabcolsep}{3.8pt}
\begin{tabular}{lcccc}
\toprule
\textbf{Model} & $\overline{|u|}$ \textbf{(mm)} & \textbf{Fold (\%)} $\downarrow$ & \textbf{Smooth. (1/mm)} $\downarrow$ & \textbf{Vort.} $\downarrow$ \\
\midrule
\rowcolor{gray!20}
  SyN (ceiling) & 2.833\,$\pm$\,0.035 & 0.200\,$\pm$\,0.015 & 0.1345\,$\pm$\,0.0011 & 0.458\,$\pm$\,0.002 \\
\midrule
  PSM  & 0.609\,$\pm$\,0.018 & \underline{0.002}\,$\pm$\,0.000 & 0.1267\,$\pm$\,0.0009 & 0.477\,$\pm$\,0.004 \\
  FLE  & 0.119\,$\pm$\,0.007 & \textbf{0.000}\,$\pm$\,0.000 & 0.1052\,$\pm$\,0.0011 & 0.479\,$\pm$\,0.003 \\
  GLIA & 0.418\,$\pm$\,0.032 & \textbf{0.000}\,$\pm$\,0.000 & 0.0893\,$\pm$\,0.0005 & 0.570\,$\pm$\,0.001 \\
\bottomrule
\end{tabular}
\end{table}

Figure~\ref{fig:qualitative} shows a qualitative displacement-magnitude comparison and Table~\ref{table:overlap_metrics} shows a quantitative analysis of the tissue overlap metrics.
All models improve the tissue overlap in the 10~mm band and over the whole brain.
PSM gives the largest improvements on both $\Delta$Dice metrics and on net reclassification at the 10~mm band.
Figure~\ref{fig:quantitative} shows quantitative comparisons between the solvers. For the tissue-specific $\Delta$Dice within the 10~mm band, PSM performs best or equal within the biophysical solvers for CSF, GM and WM and places second, behind GLIA for VT. This is expected for a solver that explicitly models ventricle compression. FLE is weakest on the overlap metrics: its forward, fixed-parameter elastic field stays sharply concentrated at the tumor rim and is too small across the bulk of peritumoral tissue to shift a tissue boundary, so it improves every tissue the least. 
Finally, SyN bounds the achievable overlap from above yet is itself far from perfect.

Regarding similarity to the SyN field, Figure~\ref{fig:quantitative} c) shows that PSM's field is most similar to SyN's. GLIA scores the lowest weighted cosine similarity, which is expected as its displacement accumulates over the whole tumor trajectory and is specifically aimed at ventricle compression.

Table~\ref{table:field_metrics} reports deformation-field characteristics that indicate physical plausibility. All biophysical models show negligible folding. SyN's folding ratio is low but two orders of magnitude larger. This does not contradict SyN's diffeomorphic construction, which holds for continuous flow, whereas $\det J$ is measured by finite differences on the 1~mm grid after composition of discrete fields. The atlas-to-patient problem is furthermore not diffeomorphic, as brain tissue is replaced with tumor tissue rather than displaced, so an image-based objective trades invertibility for overlap. Smoothness and vorticity are only weakly discriminative among the biophysical solvers.

In addition to the 134 BraTs samples that all methods were evaluated on, we additionally evaluate the PSM solver on 984 disjoint BraTS2021 samples (feasible due to fast computation times). In this evaluation, the alpha-optimization of $\alpha=1000$ from the main cohort was left constant and reproduced the per-tissue dice improvements of the main cohort. The single global force scale, therefore, transfers to unseen subjects.

\section{Conclusion}
We show that our lightweight, differentiable, energy-based model of glioma mass effect explains brain tissue deformation better than existing biophysical baselines. Our solver works on a precomputed tumor concentration and only needs a single forward solve rather than a per-patient inversion, and therefore runs orders of magnitude faster. The introduced quality metrics allow us to compare methods against each other and against a simulation without mass effect. A clear limitation is that true validation requires longitudinal preoperative studies that include healthy reference scans, so that the predicted deformation can be compared to observed tissue displacement rather than only to tissue-overlap proxies. Furthermore, a stratified analysis by brain mechanics features (such as tumor location, size, and surrounding tissue) would show where a global $\alpha$ is sufficient and where a location-specific $\alpha$ would improve the prediction accuracy of PSM. In contrast to purely image-based registration methods \cite{avants2009advanced}, our method is based on first-principles physics. This gives physical meaning to the resulting deformation field and provides an important constraint for medical applications such as radiotherapy planning. The same physical grounding also enables additional applications, such as simulating tumor resection and the resulting intraoperative brain deformation. Its differentiability also lets it be integrated into neural surrogate solvers \cite{weidner2026brain} or discrete optimization-based generative models \cite{balcerak2026graphenergymatching}, toward learning fast approximations of tumor growth and mass effect within a single gradient-based framework. Overall, our results suggest that a simple differentiable energy model can recover clinically relevant mass-effect patterns at substantially lower computational cost than established biophysical solvers. This makes the proposed solver a practical step toward fast, physics-guided deformation modeling for personalized tumor-growth simulation.

\begin{credits}
\subsubsection{\ackname} This research was funded by the German Ministry of Research, Technology and Space (BMFTR) within the Software Campus program, funding code: 16IS23069. The responsibility for the content of this publication lies with the author.

\vspace{1em}
\subsubsection{Publication notice.}
\noindent\footnotesize
This version of the contribution has been accepted for publication, after peer review (when applicable) but is not the Version of Record and does not reflect post-acceptance improvements, or any corrections. The Version of Record is available online at: (DOI to be added upon publication). Use of this Accepted Version is subject to the publisher's Accepted Manuscript terms of use \url{https://www.springernature.com/gp/open-research/policies/accepted-manuscript-terms}.
\end{credits}

\newpage
\bibliographystyle{splncs04}
\bibliography{bib}
\end{document}